\documentstyle[11pt,cal97,psfig]{article}

\begin{document}

\title{WFPC2 CTE Characterization}

\author{Andrew E. Dolphin}
\affil{National Optical Astronomy Observatories, Tucson, AZ 85726; dolphin@noao.edu}

\begin{abstract}
The limiting factor of the accuracy of WFPC2 photometry is the CTE loss, which has increased to the level of 50\% or more for faint stars at the top of the chips.  I describe recent work on characterizing this effect, and provide improved equations for CTE correction.  I also examine issues affecting background measurement, which if not done correctly can introduce artificial nonlinearities into photometry.
\end{abstract}


\section{Introduction}

Several obstacles inhibit the obtaining of accurate photometry from WFPC2 images.  The most severe of these is charge loss during readout, commonly known as CTE loss.  While no CCD is likely to be perfect, the effect is pronounced on WFPC2, where initial measurements showed that a star at the top of the chip ($y = 800$) would lose approximately 10\% of its charge while being read out.  This loss was reduced by cooling the camera from $-76^{\circ}$C to $-88^{\circ}$C, and Holtzman et al. (1995) found that the CTE loss could be corrected to acceptable levels with a correction of $0.04\ y/800$ magnitudes to photometry.  Unfortunately, this has increased over time, and faint stars at the top of the chip now can lose well over half their light to CTE loss.

Improved characterizations of the charge loss were produced by Stetson (1998), Whitmore, Heyer, \& Casertano (1999), and Saha, Labhardt, \& Prosser (2000) by analysis of larger data sets using DAOPHOT, IRAF apphot, and DoPHOT, respectively.  While there were several issues of agreement between these studies, there were also significant discrepancies.  For example, Stetson (1998) found no significant time dependence, while Whitmore et al. (1999) did.  Likewise, Saha et al. (2000) found no X-CTE loss and no nonlinearity, while both were observed in the other studies.  The fact that a different photometry package was used in each study raised the possibility that the observed CTE loss was partly a function of the package used.

Another possible source of photometric error was reported from measurements of the same fields in long and short exposures, in which objects appeared fainter in the short exposure than in the long exposure.  This was independent of the position on the chip, and the phenomenon has become known as the ``long vs. short anomaly''.  Casertano \& Mutchler (1998) characterized this effect using short and long observations of NGC 2419, and found it to be a function of counts rather than exposure time.  The effect they found was a very large one; rather than the 5\% effect reported previously, their correction is 0.18 magnitudes at 280 electrons (50 ADU at gain of 7).  However, Stetson (1998) solved for a position-independent effect as part of his CTE study and found none.  Given that his data set included the NGC 2419 observations, this only added to the questions about how well these effects were really understood.

Dolphin (2000a; hereafter D00) presented a study of the WFPC2 CTE based on reductions of 843 WFPC2 images of $\omega$ Cen and NGC 2419 using his HSTphot photometry package (Dolphin 2000b).  In addition to providing yet another CTE solution based on yet another photometry package, this work shed light on the discrepancies noted above.  The lack of time dependence seen by Stetson was observed to be primarily the result of an insufficient time baseline and the fact that all of his high-background data were obtained at the end of the time baseline.  The lack of a nonlinearity in the study of Saha et al. was found to result from their background measurements, which contained significant amounts of starlight.  This meant that their CTE correction (explicitly a function of background only) was implicitly a function of brightness as well.  Finally, it was demonstrated that the long-short anomaly is primarily a result of poor background measurements, as my reductions of the NGC 2419 data showed no significant discrepancies after the CTE loss was corrected.

In this paper, I describe ongoing efforts to improve upon D00 and present the latest results.

\section{Observations}

The basis of my CTE study (both D00 and the present work) has been the comparison of instrumental magnitudes of WFPC2 images with ground-based photometry.  This allows the use of the ground-based data as the standard stars, and any discrepancy is understood as resulting from a combination of CTE loss and calibration.  This is not the only way to do such a work; Whitmore et al. (1999) based their study on relative photometry of stars as they were moved around the chips.  Both techniques have their drawbacks; the main drawback of the route I chose is that errors in the functional form can be harder to find.  Most notably, a position-independent nonlinearity such as the long-short anomaly could be fit out as CTE loss.  I address this concern in the next section.

In D00, I used images of the $\omega$ Cen standard field and of NGC 2419.  The mixture of two fields was necessary because, at the time, no images of the $\omega$ Cen standard field had high background levels; however this introduced the possibility of errors in the CTE correction caused by inconsistent calibrations of the two ground-based data sets.  This compromise is no longer necessary, as high-background images of $\omega$ Cen have been taken, and thus in this work I use only those data.  The ground-based photometry is that of Walker (1994), which I have transformed to the expected WFPC2 flight system magnitudes by using the Holtzman et al. (1995) transformations.

A total of 1216 images were photometered in this project, covering all observations of the $\omega$ Cen field in $B,V,R,I$ filters through August 2002.  The majority of the images (800) were taken in the F555W and F814W filters.  All images were photometered with HSTphot.  Because of the huge number of stars observed, I eliminated all data I considered suspect because the aperture corrections were unusual or were based on too few stars.

After matching the stars on the WFPC2 images to the list of ground-based standards, many of the standards were eliminated because they were resolved into multiple stars or small extended objects by WFPC2.  A few other standards were deemed to be poorly photometered because no CTE correction was capable of producing photometry that agreed at the 10\% level; these were also eliminated from the sample.

The end result was a list of 36983 stars on the WFPC2 images that had been matched to any of 202 of Walker's standard stars.  From this list, the magnitude differences between the WFPC2 magnitudes and ground-based magnitudes were fit as functions of CTE loss and zero point differences.  Since this contribution describes CTE results, the zero points will not be discussed further.  They are, however, available from the author's web site.

\section{Characterization}

It is important to bear in mind that the correction formulae below are based on the functional forms that best fit the data, and are not based on a physical understanding of the charge transfer process.  It is likely that a ``perfect'' correction would be much more complex; what is presented below is the best fit using a minimum number of parameters.

\begin{figure}
\hspace{1.3in}
\psfig{file=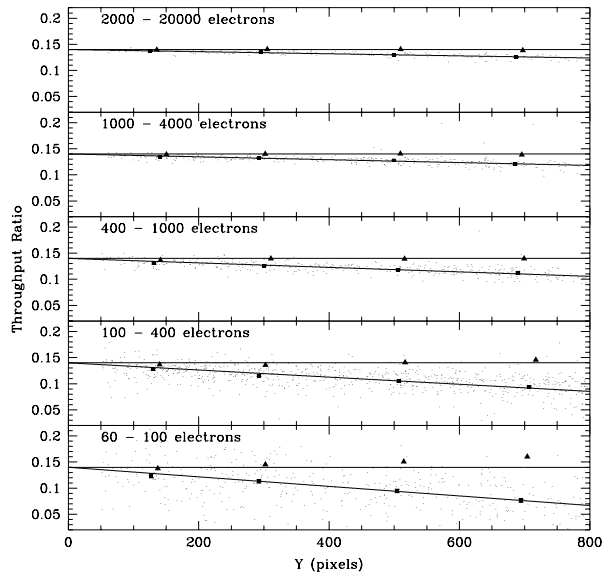,height=3.0in}
\caption{Ratio of counts in short (14 second) image to counts in long (100 second) image, separated by star brightness in the short image.  Small dots are measurements of individual stars; the square and diagonal line are the fit to the trend.  The triangles show the data after CTE correction using old correction formulae, while the horizontal line shows the expected ratio of 0.14.  Note that the bottom panel shows a significant overcorrection, indicated by the triangles falling well above the horizontal line.}
\label{fig_cteold}
\end{figure}

\begin{figure}
\hspace{1.3in}
\psfig{file=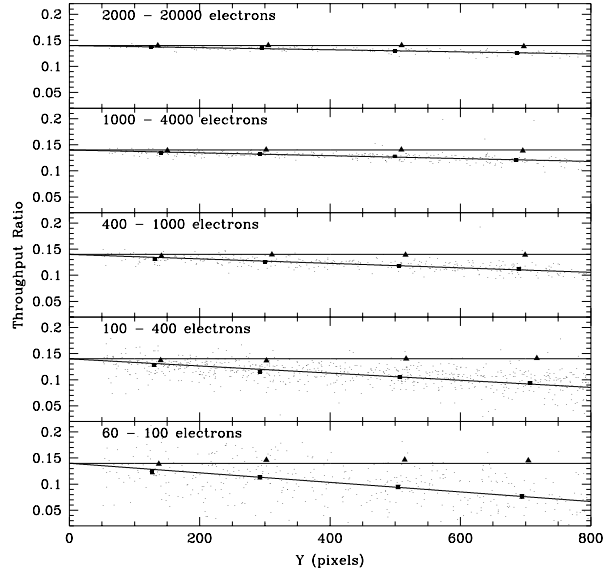,height=3.0in}
\caption{Same as Figure \ref{fig_cteold}, but using the new CTE corrections.  Note that the bright stars are still fit well, while the overcorrection for faint stars is eliminated.}
\label{fig_ctenew}
\end{figure}

A feature of the CTE correction procedure is that it includes the effect of CTE loss on the CTE loss.  That is, as a star reads out and becomes fainter, CTE loss (in magnitudes per pixel) increases.  This leads to a more complex functional form, but gives a correction that is accurate down to about 60 electrons ($< 10$ ADU at gain 7) instead of only to about 100 electrons.  At much fainter levels ($20-30$ electrons), it is clear that the CTE loss is less than what is predicted, since otherwise noise peaks in the background level itself (which can be thought of as faint stars) would be truncated.  The improvement can be seen in Figures \ref{fig_cteold} and \ref{fig_ctenew}, which show count ratios between observations in short and long images as a function of $y$ position and counts.  It is clear that the corrections above 100 electrons work well either way, but that the newer correction (Figure \ref{fig_ctenew}) is much better for stars between 60 and 100 electrons.

It should also be noted that the ground-based standard magnitudes were not used in creating Figures \ref{fig_cteold} and \ref{fig_ctenew}; rather they are based entirely on instrumental WFPC2 magnitudes.  It is clear that the corrections performed well, dispelling any concern that the use of absolute photometry comparisons rather than relative photometry comparisons impaired the solution.  Most notably, had a long-short anomaly been present, the $y$ intercept of the uncorrected photometry (the diagonal line in each panel) would have been lower than 0.14, while the CTE correction would have overcorrected stars with high $y$ values and undercorrected those with low $y$ values since the long-short anomaly would have been fit by CTE loss.  We can clearly see from the figure that this is not the case.  If long-short anomaly were at the level seen by Casertano \& Mutchler (1998), the $y$ intercept would have been at 0.11 in the $100-400 e^{-}$ panel and at 0.13 in the $400-1000 e^{-}$ panel.  (A comparison with the bottom panel is unwarranted, as their correction was only valid to 200 electrons.)  Thus it is clear that any long-short error is at least an order of magnitude smaller than what they reported.

\section{CTE Correction Recipe}

The procedure for correcting for CTE loss is outlined below.  The XCTE correction depends only on $x$ and the background (in electrons).  Note that the background value used should be the true background at the position of the star, rather than the background measured nearby a star (which contains some starlight).  In images with variable background, this requires a knowledge of the amount of starlight contained in the background measurement.

\begin{equation}
bg \equiv \sqrt{1 + \hbox{background}^2} - 10
\end{equation}
\begin{equation}
\hbox{XCTE} = 0.0194 e^{-0.00085 bg} x/800
\end{equation}

The YCTE loss depends on $y$, background, brightness (also in electrons), and the date of the observation.

\begin{equation}
lbg \equiv 0.5 \ln(1 + \hbox{background}^2) - 1
\end{equation}
\begin{equation}
lct \equiv \ln(\hbox{brightness}) + 0.921 \hbox{XCTE} - 7
\end{equation}
\begin{equation}
yr \equiv (\hbox{MJD}-50193) / 365.25
\end{equation}
\begin{equation}
c_1 \equiv 0.0143 ( 0.729 e^{-0.397 lbg} + 0.271 e^{-0.0144 bg} ) ( 1 + 0.267 yr - 0.0004 yr^2 ) y/800
\end{equation}
\begin{equation}
c_2 \equiv 2.99 e^{-0.479 lct}
\end{equation}
\begin{equation}
\hbox{YCTE} = \ln [ (1+c_2) e^{c_1} - c_2 ] / 0.441 
\end{equation}

Both XCTE and YCTE corrections are in magnitudes, and should be subtracted from instrumental magnitudes to make the correction.

Figures \ref{fig_CTEy}, \ref{fig_CTEx}, \ref{fig_CTEbg}, and \ref{fig_CTEct} show the magnitude differences between the WFPC2 magnitudes and the ground-based standard magnitudes before and after correction.  In all figures, the top panel shows the uncorrected WFPC2 minus ground-based magnitude and the bottom panel shows the corrected difference.

\begin{figure}
\vspace{2.5in}
\caption{WFPC2 magnitude errors (observed minus ground-based standard magnitudes) vs. $y$, before (top) and after (bottom) corrections were applied.  Note that stars with large $y$ values have fainter raw magnitudes, but no discernible trend remains in the CTE-corrected magnitudes.}
\label{fig_CTEy}
\end{figure}

\begin{figure}
\vspace{2.5in}
\caption{Like Figure \ref{fig_CTEy}, plotted vs. $x$.  Note that stars with large $x$ values have slightly fainter raw magnitudes, but no discernible trend remains in the CTE-corrected magnitudes.}
\label{fig_CTEx}
\end{figure}

\begin{figure}
\vspace{2.5in}
\caption{Like Figure \ref{fig_CTEy}, plotted vs. background.  Note that stars on low background have fainter raw magnitudes, but no discernible trend remains in the CTE-corrected magnitudes.}
\label{fig_CTEbg}
\end{figure}

\begin{figure}
\vspace{2.5in}
\caption{Like Figure \ref{fig_CTEy}, plotted vs. brightness.  Note that faint stars have fainter raw magnitudes, but no discernible trend remains in the CTE-corrected magnitudes.}
\label{fig_CTEct}
\end{figure}

\section{Background Measurement}

Although this work focuses on CTE corrections, another critical issue in obtaining accurate photometry is background measurement.  If the background is mismeasured, one will introduce a nonlinearity into the photometry.  In fact, it is likely that the Casertano \& Mutchler (1998) study of the long vs. short anomaly was influenced by background determination.  Two pieces of evidence point to this.  First, their residuals are better fit by such a function than by their reported correction formula.  Second, their discrepancies were larger when using larger photometry apertures.  In addition, Hill et al. (1998) noted that the short vs. long error they measured could be explained by a $2 e^{-}$ per pixel loss only in pixels containing stars, which is the same as a $2 e^{-}$ per pixel overestimation of the sky.

\begin{figure}
\hspace{0.5in}
\psfig{file=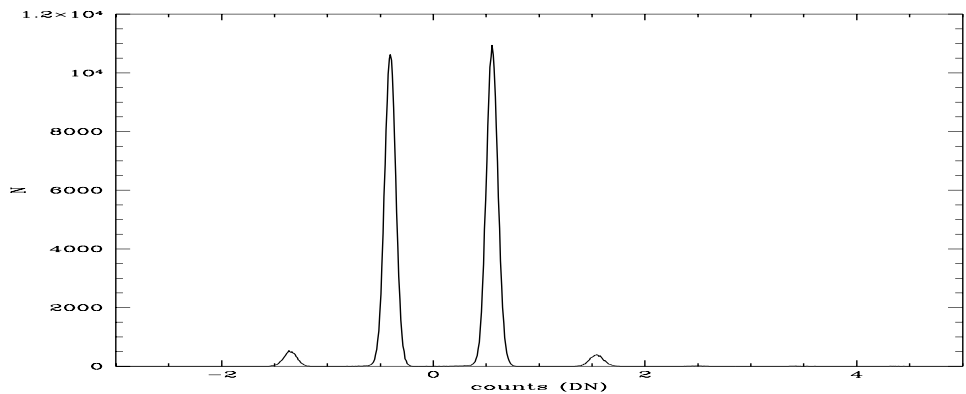,height=2.0in,width=5.0in}
\caption{Histogram of PC data values from a short (14 second) image.}
\label{fig_skyhist}
\end{figure}

The reason special care must be taken in calculating background levels is the very low background levels of WFPC2 observations.  The commonly-used IRAF packages were designed for reducing ground-based data, for which the histogram of background pixel levels can be approximated as a Gaussian or other smooth function.  However, when the noise in the background is less than 1 ADU, the histogram is dominated by digitization.  A sample histogram of sky values from a 14 second exposure is shown in Figure \ref{fig_skyhist}.

\begin{figure}

\hspace{0.7in}
\psfig{file=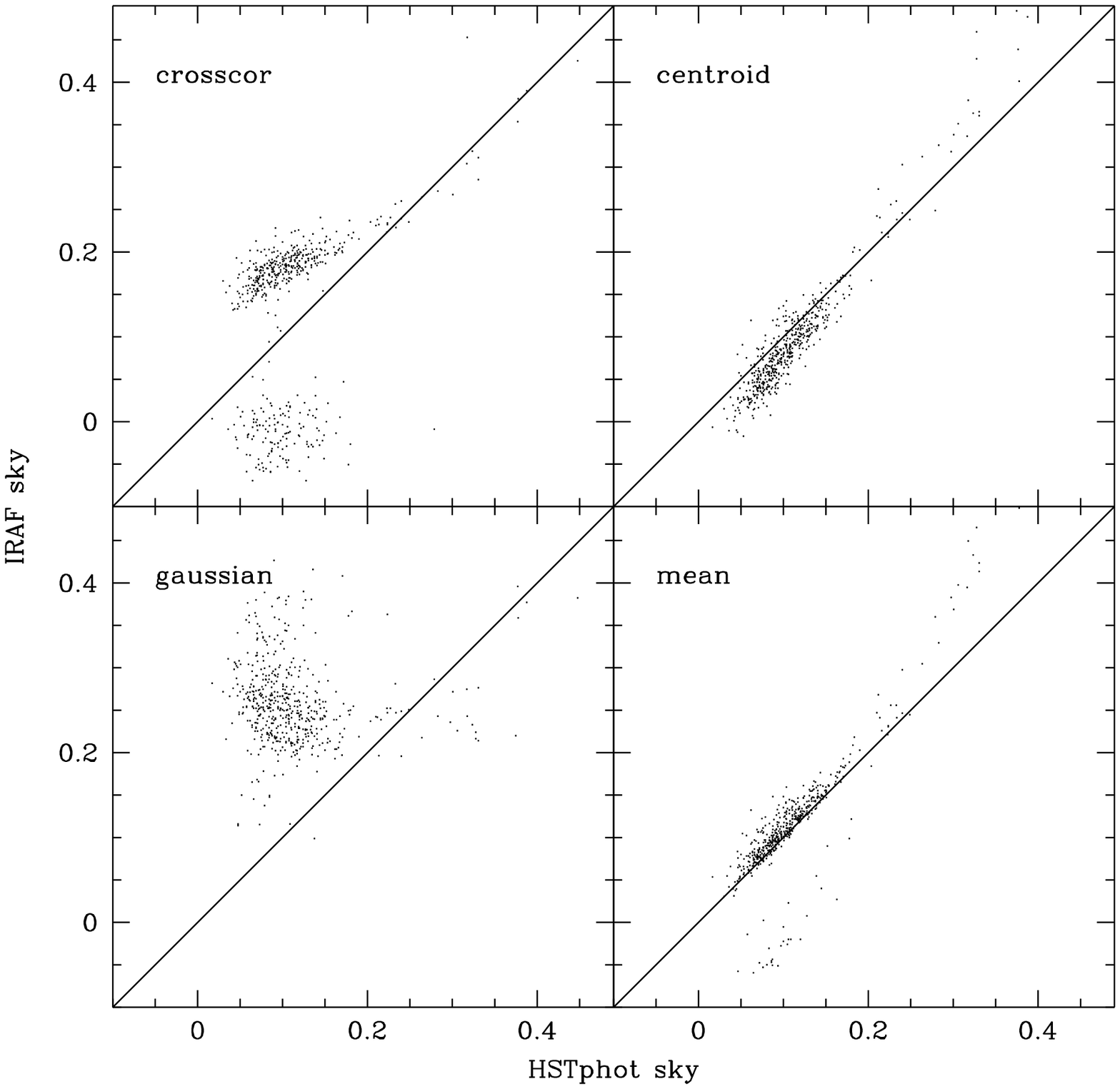,height=3.5in}

\vspace{24pt}
\hspace{0.7in}
\psfig{file=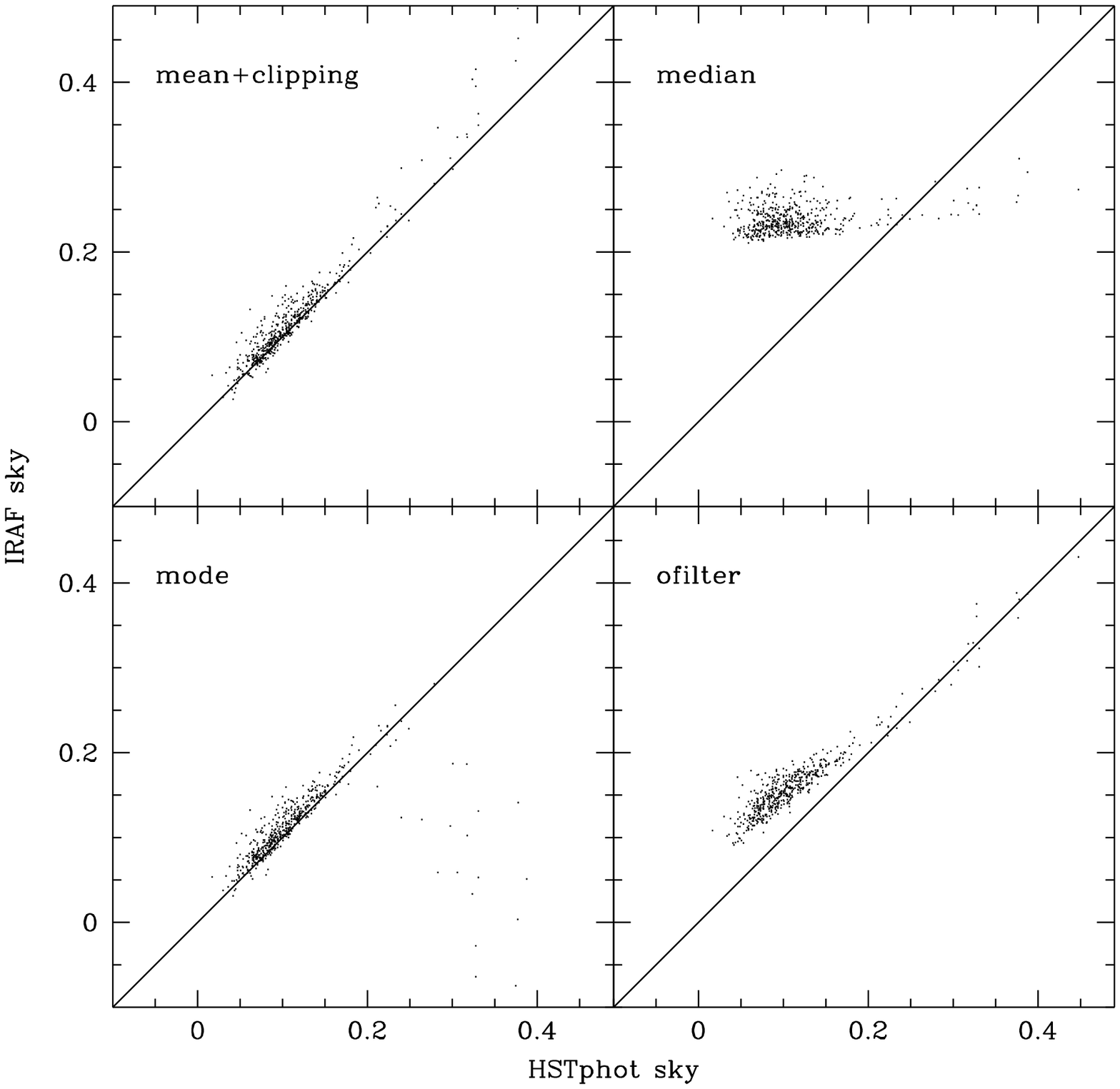,height=3.5in}

\caption{Sky determinations from various IRAF algorithms.}
\label{fig_skycomp}
\end{figure}

Naturally any sky-measuring algorithm that assumes a smooth distribution is prone to failure when handling these data.  Figure \ref{fig_skycomp} show measurements made by various IRAF sky algorithms on this image, with most of the default parameters left in place.  This analysis is similar to the examination done by Ferguson (1996) on simulated data.  In all panels, the $x$ values of the sky are taken from HSTphot measurements, which use a sigma-clipped mean algorithm that has been verified to work in cases like these.

From a comparison of the plots, it is clear that not all of the algorithms function as intended when facing digitization-dominated sky histograms.  The Gaussian and median routines appear to be affected by the histogram peak at $~\sim 0.6$ ADU, the cross-correlation algorithm is extremely unstable, and the optimal filtering algorithm is biased.  The ``mode'' calculation is not a true mode, but rather a calculation including the median routine and is thus not suitable.  Additional tests have been made involving simulated images with known background levels, and these routines handle the simulated data no better than the real data.

In more detailed tests, it is observed the centroid algorithm is also prone to small biases, while a straight mean will be overly affected by deviant pixels (uncorrected bad pixels or faint stars).  The errors in the centroid algorithm affect measurements at only the 0.03 ADU level, however, which is acceptable for most purposes.  Likewise, the stability of the mean calculation can be increased by use of sigma clipping.  Thus the recommendation is that either routine can be used successfully, but that a sigma-clipped mean is preferable.

\section{Ongoing Work}

While the current CTE corrections work very well down to extremely faint count levels, there are two key problems yet to be addressed in a comprehensive manner.  First is the issue of extremely faint stars.  An examination of the CTE correction equations shows that the expected CTE loss becomes very large for faint stars.  Specifically, the noise peaks in the background itself should be destroyed by CTE loss, as those peaks should be treated as 1 or 2 ADU stars.  The histogram in Figure \ref{fig_skyhist} indicates that CTE does affect the background but not at the level predicted by the CTE correction equations.  The dilemma is that it is prohibitively difficult to obtain accurate photometry of stars of brightness 4 ADU (28 electrons at gain of 7).  Recent calibration data have been obtained that should allow this issue to be addressed, however, with repeated short exposures of NGC 2419 that can be coadded to improve the photometry.

A second problem affecting photometry is that the star's profile is not dimmed uniformly.  In a fractional sense, the highest charge loss comes from the bottom edge and sides of the star, while the top edge can actually have charge added.  Furthermore, the ratio of dimming of the wings to the dimming of the peak will be a function of the star brightness.  An example of this is shown in Figure \ref{fig_sub}, which shows the difference between a short and long image, scaled so that they should cancel out.  The trail above the star in the image is the light that was lost by the star and released from the trap several readout steps later.

\begin{figure}
\vspace{2.5in}
\caption{Smoothed difference image (short $-$ scaled long image), showing charge loss (dark) at star's position and trail (bright) above the star.  The brightness of the trail is $\sim 0.1$ counts.}
\label{fig_sub}
\end{figure}

The result of this is that PSF-fitting photometry will have some trouble dealing with faint stars, which will have different PSFs from the bright stars.  Furthermore, the possibility exists for small systematic differences between PSF-fitting and aperture photometry.  While both errors are dwarfed by the random error in measuring faint stars, there are some applications for which they are significant.  More significantly, the effects of CTE loss on the profiles of extended objects can be quite large.  Riess (2000) created a simple model for trapping and recreate the CTE effects on the profiles of extended sources.  The challenge will be to improve such a model so that it can quantitatively reproduce the CTE loss measured in stellar photometry.  Given the complexity of the correction equations, it is clear that this is not a trivial task.  However, once this is achieved, one can invert the charge loss model to obtain the true image prior to readout, effectively correcting the image for CTE loss rather than the photometry.

\section{Summary}

I have developed and described a CTE correction procedure to supersede that of D00.  There are several significant improvements over the previous corrections.  First, a single uniform set of ground-based photometry is used to provide the comparisons.  Second, HSTphot has been improved several times over the intervening time.  Third, more stringent cuts were applied to the data to eliminate bad points.  Finally, an improved functional form of the CTE correction accounts for the changing CTE loss as a star becomes fainter during readout.

I have also examined a commonly-overlooked aspect of stellar photometry, background measurement.  In short images, the background value is sufficiently low that the digitization effects dominate the histogram.  If not handled correctly, this can result in significant artificial nonlinearities added to the data.  After exploring the options offered by IRAF, the recommendation is that one use the ``mean'' algorithm with sigma clipping.

Finally, I conclude by describing ongoing efforts to improve the CTE corrections for faint sources and to obtain an image-based CTE correction.  My WFPC2 calibration web site (http://www.noao.edu/staff/dolphin/wfpc2\_calib/) will be kept updated with improvements to the CTE corrections and photometric zero points as they become available.

%\acknowledgments

\end{document}